\documentclass[aps,prl,preprint,superscriptaddress,showpacs]{revtex4}
\usepackage{dcolumn}
\usepackage{amsfonts}
\usepackage{amssymb}
\usepackage{amsmath}
\usepackage{graphicx}

\usepackage{dcolumn}   
\usepackage{bm}        

\begin{document}

\title{Two-Dimensional Confinement of 3$d^1$ Electrons in LaTiO$_3$/LaAlO$_3$
Multilayers}

\date{\today}

\author{S. S. A. Seo}\email{seos@ornl.gov}
\altaffiliation[Current address: ]{Materials Science and
Technology Division, Oak Ridge National Laboratory, Oak Ridge, TN
37831, USA} \affiliation{Department of Physics and Astronomy,
Seoul National University, Seoul 151-747, Korea}
\affiliation{Max-Planck-Institut f\"{u}r Festk\"{o}rperforschung,
Stuttgart D-70569, Germany}
\author{M. J. Han}
\altaffiliation[Current address: ]{Department of Physics, Columbia
University, New York, NY 10027, USA} \affiliation{Department of
Physics and Astronomy, Seoul National University, Seoul 151-747,
Korea}
\author{G. W. J. Hassink}
\affiliation{Department of Advanced Materials Science, University
of Tokyo, Kashiwa, Chiba 277-8561, Japan} \affiliation{MESA+
Institute for Nanotechnology, University of Twente, Enschede
NL-7500 AE, The Netherlands}
\author{W. S. Choi}
\affiliation{Department of Physics and Astronomy, Seoul National
University, Seoul 151-747, Korea}
\author{S. J. Moon}
\affiliation{Department of Physics and Astronomy, Seoul National
University, Seoul 151-747, Korea}
\author{J. S. Kim}
\altaffiliation[Current address: ]{Department of Physics, Pohang
University of Science and Technology, Pohang, Kyungbuk 790-784,
Korea}\affiliation{Max-Planck-Institut f\"{u}r
Festk\"{o}rperforschung, Stuttgart D-70569, Germany}
\author{T. Susaki}
\affiliation{Department of Advanced Materials Science, University
of Tokyo, Kashiwa, Chiba 277-8561, Japan}
\author{Y. S. Lee}
\affiliation{Department of Physics, Soongsil University, Seoul
156-743, Korea}
\author{J. Yu}
\affiliation{Department of Physics and Astronomy, Seoul National
University, Seoul 151-747, Korea}
\author{C. Bernhard}
\affiliation{Department of Physics and Fribourg Center for
Nanomaterials, University of Fribourg, 1700 Fribourg, Switzerland}
\author{H. Y. Hwang}
\affiliation{Department of Advanced Materials Science, University
of Tokyo, Kashiwa, Chiba 277-8561, Japan} \affiliation{Japan
Science and Technology Agency, Kawaguchi 332-0012, Japan}
\author{G. Rijnders}
\affiliation{MESA+ Institute for Nanotechnology, University of
Twente, Enschede NL-7500 AE, The Netherlands}
\author{D. H. A. Blank}
\affiliation{MESA+ Institute for Nanotechnology, University of
Twente, Enschede NL-7500 AE, The Netherlands}
\author{B. Keimer}
\affiliation{Max-Planck-Institut f\"{u}r Festk\"{o}rperforschung,
Stuttgart D-70569, Germany}
\author{T. W. Noh}\email{twnoh@snu.ac.kr}
\affiliation{Department of Physics and Astronomy, Seoul National
University, Seoul 151-747, Korea}

\begin{abstract}

We report spectroscopic ellipsometry measurements of the
anisotropy of the interband transitions parallel and perpendicular
to the planes of (LaTiO$_3$)$n$(LaAlO$_3$)5 multilayers with $n$ =
1$-$3. These provide direct information about the electronic
structure of the two-dimensional (2D) 3$d^1$ state of the Ti ions.
In combination with LDA+$U$ calculations, we suggest that 2D
confinement in the TiO$_2$ slabs lifts the degeneracy of the
$t_{2g}$ states leaving only the planar $d_{xy}$ orbitals
occupied. We outline that these multilayers can serve as a model
system for the study of the $t_{2g}$ 2D Hubbard model.
\end{abstract}
\smallskip


\maketitle

Recent advances of oxide thin-film synthesis techniques enable the
study of oxide multilayers with atomically abrupt interfaces
\cite{Ohtomo_nature, HNLee}. Pioneering studies on various oxide
heterostructures have revealed intriguing physical phenomena such
as electronic reconstruction \cite{Millis, Seo}, quantum Hall
effect \cite{Tsukazaki}, and orbital reconstruction \cite{Chak} at
the interfaces. A new approach of dimensionality-control of oxides
also has been made possible by the potential well (or quantum
well) geometry of La$M$O$_3$/LaAlO$_3$ ($M$: transition metal
elements) since the Al 3$p$ state is located much higher in energy
than the transition metal 3$d$ state \cite{Hotta}. Recent
theoretical studies have brought particular attention to the
potential well geometry since intriguing physical properties can
be manipulated in such multilayered structures. For instance,
high-$T_c$ superconductivity was predicted to occur in
LaNiO$_3$/LaAlO$_3$ \cite{Jiri}.

In this letter, we report the electronic structure and orbital
reconstruction of 3$d^1$ electrons in multilayers consisting of a
few unit-cell LaTiO$_3$ (LTO) layers embedded in LaAlO$_3$ (LAO)
using optical spectroscopic ellipsometry and LDA+$U$ calculations
(LDA: local density approximation). Single crystalline LTO is a
Mott insulator with a small Mott-Hubbard gap ($\Delta_{MH}$) of
$\sim$0.2 eV \cite{Okimoto} while LAO is a band insulator with a
wide bandgap of $\sim$5.6 eV. By taking into account the
electronic structures of the bulk phases, a two-dimensional (2D)
confinement of the Ti 3$d^1$ state in LTO/LAO multilayers can be
considered (Fig. 1 (a)) similarly to the V 3$d^2$ state in
LaVO$_3$/LAO of Ref. \cite{Hotta}. This 2D Ti$^{3+}$ state is
particularly interesting since a bulk material possessing a
Ti$^{3+}$O$_2$ 2D square-lattice has not yet been found. In
analyzing the interband optical transitions, we show that a 2D
confined 3$d^1$ Mott state can be realized in the LTO/LAO
multilayers. Along with the confinement, the Ti 3$d$ orbitals are
also reconstructed as the three-fold degeneracy of the $t_{2g}$
levels is lifted yielding partially occupied $d_{xy}$- and empty
$d_{yz,zx}$-orbitals.

By using the pulsed laser deposition technique, we grew
multilayers of ((LTO)$n$(LAO)5)$\times$20, which means twenty
repetitions of $n$ (=1, 2, and 3) pseudo-cubic perovskite
unit-cell(s) of LTO ($\sim$ $n\times$3.96 \AA) and five
pseudo-cubic unit-cells of LAO ($\sim$ 5$\times$3.78 \AA), on
SrTiO$_3$ (STO) (001) substrates. (Details about the growth can be
found in the supplementary online material \cite{Supplementary}.)
The relevant parameter concerning quantum confinement is the ratio
of the potential well width of the multilayers and the excitonic
radius $a_0=a_{B}\varepsilon/m^{*}$ \cite{Cox}, where
$a_{B}$(=0.53 \AA) is the hydrogenic Bohr radius, $\varepsilon$ is
the dielectric permittivity, and $m^{*}$ is the effective
electronic mass in LTO. With reasonable estimates of
$m^{*}\approx$ 2$-$4 and $\varepsilon\approx$ 20$-$50, we obtain
$a_0 \approx$ 3$-$13 \AA. Hence, in this study we pursued LTO
layers with $n<4$.

Figure 1(b) shows x-ray $\theta$-2$\theta$ scans around the STO
002-reflection, which reveal sharp superlattice satellite peaks
due to the periodicity of the multilayer. The $\Delta$$l$ between
the satellite peaks satisfies the relation $\Delta$$l$=1/($n$+5)
in each (LTO)$n$(LAO)5 multilayer. X-ray reciprocal space mappings
confirmed that the averaged in-plane lattice constants were
coherently strained to those of the STO substrates. Although a
non-stoichiometric phase with excessive oxygen LaTiO$_{3+\delta}$
\cite{Schmehl} or La vacancies La$_{1-x}$TiO$_3$ \cite{Crandles}
is known to be metallic, our multilayers were highly insulating in
the measurements of $dc$-conductivity and optical absorption
spectroscopy \cite{Seo_APL}.

To investigate the electronic structure of the (LTO)$n$(LAO)5
multilayers, we used bulk-sensitive spectroscopic ellipsometry in
the ultraviolet (UV) photon energy region, i.e. 3.3$-$6.5 eV,
which is compatible with the energies of interband optical
transitions of LTO. Spectroscopic ellipsometry is a
self-normalizing technique that directly measures the complex
dielectric function
$\tilde{\varepsilon}$($\omega$)[=\emph{$\varepsilon$}$_1$(\emph{$\omega$})+\emph{$i\varepsilon$}$_2$(\emph{$\omega$})]
of a multilayer without the need of Kramers-Kronig transformation.
(See the supplementary material for details on the spectroscopic
ellipsometry measurements and analyses \cite{Supplementary}.)
Since the probing depth of this technique is typically longer than
about 500 \AA, it is very useful to characterize buried interfaces
and layers. Ellipsometry is also advantageous in determining the
in-plane and out-of-plane optical responses of anisotropic
materials.

Figures 2 shows the anisotropy of the optical conductivity spectra
($\sigma_{1}$($\omega$)) of the (LTO)$n$(LAO)5 ($n$=1, 2, and 3)
multilayers. They were obtained by using the relation of
$\tilde{\varepsilon}(\omega)=\varepsilon_1(\omega)+i4\pi\sigma_1(\omega)/\omega$.
The parameters characterizing the optical transitions were
obtained by fitting to Lorentz oscillators:
\begin{equation}
\tilde{\varepsilon}(\omega)=\epsilon_\infty+\sum_j\frac{S_{0j}\cdot\omega^{2}_{0j}}{\omega^{2}_{0j}-\omega^{2}-i\omega\Gamma}.
\label{eq:eq1}
\end{equation}
The results of this fit procedure are shown by the solid lines.
(The values of the fitting parameters are listed in Table 1 of the
supplementary material \cite{Supplementary}.) There are two broad
peaks in the in-plane ($E//ab$) optical spectra. The low energy
peak ($\alpha$) and the high energy peak ($\beta$) can be assigned
as charge transfer transitions from the O 2$p$ state to the
unoccupied Ti 3$d$ $t_{2g}$ and to the Ti 3$d$ $e_{g}$ states,
respectively. The energy difference between the two optical
transitions gives the crystal field splitting, 10$Dq$ of the Ti
3$d$ state \cite{O-La}. It is noteworthy that the $\alpha$-peak
position increases as the thickness of LTO layers decreases while
the $\beta$-peak position remains almost unchanged. The inset of
Fig. 2(c) shows how 10$Dq$ of the Ti 3$d$ levels depends on the
LTO sublayer-thickness in comparison to the value of 1.67 eV
\cite{Higuchi} in bulk LTO.

Another notable feature is a sharp peak ($\bullet$) around 3.7 eV
in the out-of-plane ($E//c$) spectra, which has not been observed
in any bulk crystals nor thin films of LTO and LAO. In general, an
interband optical transition intensity $I_{i\rightarrow{f}}$ from
an initial state $i$ to a final state $f$ at $\hbar\omega_{0}$ can
be described as $I_{i\rightarrow{f}}(\hbar\omega_0)=\int|\langle
f|M|i\rangle|^2\rho_f(\omega)\rho_i(\omega-\hbar\omega_0)d\omega$
according to the Fermi golden rule, where $M$ is the matrix
element, and $\rho_i$ and $\rho_f$ are the densities of states for
$i$ and $f$, respectively. Hence, a sharp optical conductivity
peak usually appears when it involves both narrow-bandwidth
initial (occupied) and final (unoccupied) states such as quantized
levels in a quantum-well. Since the Ti--O hybridization becomes
weaker along the out-of-plane direction than that along the
in-plane directions, a narrowing of the bonding state and a
reduced bonding-antibonding separation are expected. Although the
origin of the peak around 3.7 eV still remains unclear at this
moment, it may be a signature of the asymmetric hybridization in
the layered structure, which causes major modifications of the
electronic structure and optical properties. Note that all these
experimental spectra cannot be explained by the 2D effective
medium approximation \cite{Agranovich} using the spectra of bulk
LTO (Ref. \cite{Arima}) and LAO (Ref. \cite{SGLim}). This also
suggests that the electronic structure of these multilayers is not
a simple average of the two mother compounds, but rather strongly
reconstructed.

To examine more details of the electronic structure and magnetic
properties, we performed LDA+$U$ calculations with the on-site
Coulomb energy $U$ ($\equiv \tilde{U}-J =$ 6 eV) \cite{MJHan,
Supplementary}, which is consistent with previous studies on bulk
LTO \cite{Igor, Pavarini}. The magnetic ground state is a
checker-board type antiferromagnetic (AFM) spin order which is
more stable than the striped AFM and ferromagnetic ordering. It is
noteworthy that the spin structure is similar to that of the
undoped cuprates. Figure 3 shows the spin-averaged partial density
of states (PDOS) of the TiO$_2$ slabs in the (LTO)$n$(LAO)5
multilayers. One of the most remarkable points is that the
three-fold degeneracy in the $t_{2g}$ state is lifted and the
$d_{xy}$-orbital is partially occupied while $d_{yz,zx}$-states
are pushed to higher energies.
On the other hand, the isotropic spin wave spectrum observed below
the N\'eel temperature (T$_N$ = 140-150 K) of bulk LTO has
suggested strong orbital fluctuations \cite{Keimer}. Such a
disordered orbital state in a cubic lattice is very unusual, and
similar disorder occurs when mobile carriers are present in $e_g$
orbital systems such as the manganites \cite{Ishihara}. Moreover,
ferromagnetic ordering is more favored than AFM ordering when
$t_{2g}$ orbitals are degenerate in the cubic lattice
\cite{Maekawa}. The theoretical description of the orbital state
of bulk LTO is still under debate (see e.g. Refs.
\cite{Orbitals}). However, in the 2D (LTO)$n$(LAO)5 multilayers,
the orbital degeneracy can be easily lifted, and a
$d_{xy}$-orbital configuration, which is different from that of 3D
bulk LTO, can be formed. This notable difference of the electronic
structure from that of the bulk counterpart is most likely caused
by the heterointerfaces. Due to the existence of the LAO layers,
the hybridization of Ti-$3d$ levels with O-$2p$ becomes
asymmetric: the $d_{xy}$ states hybridizes two-dimensionally with
the in-plane oxygens while the hybridization between $d_{yz,zx}$
states and out-of-plane oxygens is weaker. A similar planar
orbital reconstruction has also been suggested for LTO-STO 2D
superlattices \cite{Pickett}.

The lifted degeneracy of the $t_{2g}$ state and the
$d_{xy}$-orbital occupation in the (LTO)$n$(LAO)5 multilayer is
indeed consistent with the optical spectra. We estimate that the
gap energy between the $d_{xy}$-orbital and $d_{yz,zx}$-states
increases by about 0.6 eV as $n$ decreases from $n=3$ to $n=1$
(Fig. 3). Since the Ti $e_g$ states remain at the same energy, the
value of 10$Dq$ decreases as $n$ decreases. Such a change of
10$Dq$ might be induced by a local lattice distortion, i.e. a
change of the Ti-O-Ti bond angle and/or distance, which is
proportional to $d_r^{1.5}/d_{Ti-O}^{3.5}$ \cite{Harrison}, where
$d_r$ is the radial size of the $d$-orbital and $d_{Ti-O}$ is the
distance between Ti and O ions. However, in our multilayer
samples, the in-plane lattice constants are fully strained to the
STO substrates such that the lateral Ti-O distance does not change
very much. We might still have to consider stronger lattice
distortions around TiO$_6$ octahedra by local strains for the
thinner LTO layers, but the experimentally observed 10$Dq$ values
do not increase but decrease as $n$ decreases (Fig. 2(c) inset).
Hence, lattice-distortions cannot be a reason for the electronic
changes. Based on our optical spectroscopic results and LDA+$U$
calculations, Ti 3$d$ energies in the (LTO)$n$(LAO)5 multilayer
potential wells can be schematically summarized as in Figs. 4
(b)-(d). We believe that the electronic 2D confinement in the
TiO$_2$ plane plays the most important role in the stabilization
of the planar $d_{xy}$-orbital configuration.

In conclusion, our experimental and theoretical results suggest
that the Ti 3$d^1$ states have a ferro-orbital configuration with
only $d_{xy}$-orbital occupation in the 2D TiO$_2$ slabs. As we
narrow the thickness of the confined TiO$_2$ slabs to a
mono-layer, the $d_{xy}$-orbitals become more stable due to the
larger energy gap between the states of $d_{xy}$ and $d_{yz,zx}$
in a $t_{2g}$ level. The 2D confinement of a single electron in a
Ti 3$d_{xy}$ level results in an electronic structure that is
isomorphic to that of the undoped precursors of the cuprate
high-$T_c$ superconductors. It should therefore be interesting to
systematically vary the layer thickness, layer sequence, and
doping level of these structures.

SSAS thank A. Fujimori, G. Jackeli, G. Khaliullin, S. Okamoto, and
A. V. Boris for useful discussions. MJH is indebted to A. J.
Millis and C. A. Marianetti for valuable insights. This research
was supported by the Basic Science Research Program through the
National Research Foundation on Korea (NRF) funded by the Ministry
of Education, Science and Technology in Korea (No. 2009-0080567),
a nanotechnology program of the Dutch Ministry of Economic Affairs
(NanoNed), the Swiss National Science Foundation (SNF project
20020-119784), and the German Science Foundation (SFB/TRR 80).

\begin{figure}
\includegraphics[width=3.2 in, bb=106 363 460 590]{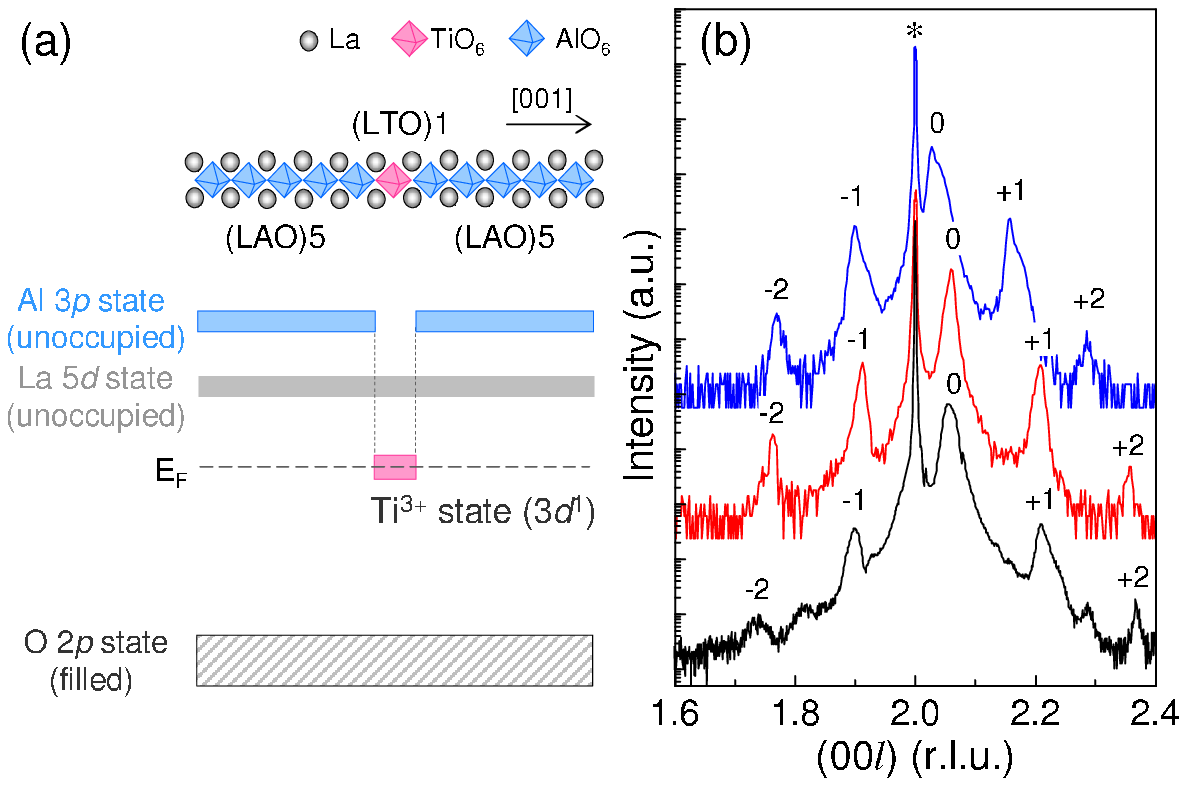}
\caption{\label{fig1} (color online) (a) Schematic of the energy
levels of a multilayer of (LTO)1(LAO)5, showing a potential well
geometry. (b) X-ray Bragg reflections around the STO 002 ($\ast$),
and superlattice satellite peaks of the (LTO)3(LAO)5,
(LTO)2(LAO)5, and (LTO)1(LAO)5 multilayers (from top to bottom).}
\end {figure}

\begin{figure}
\includegraphics[width=3.1 in,bb=137 254 446 608]{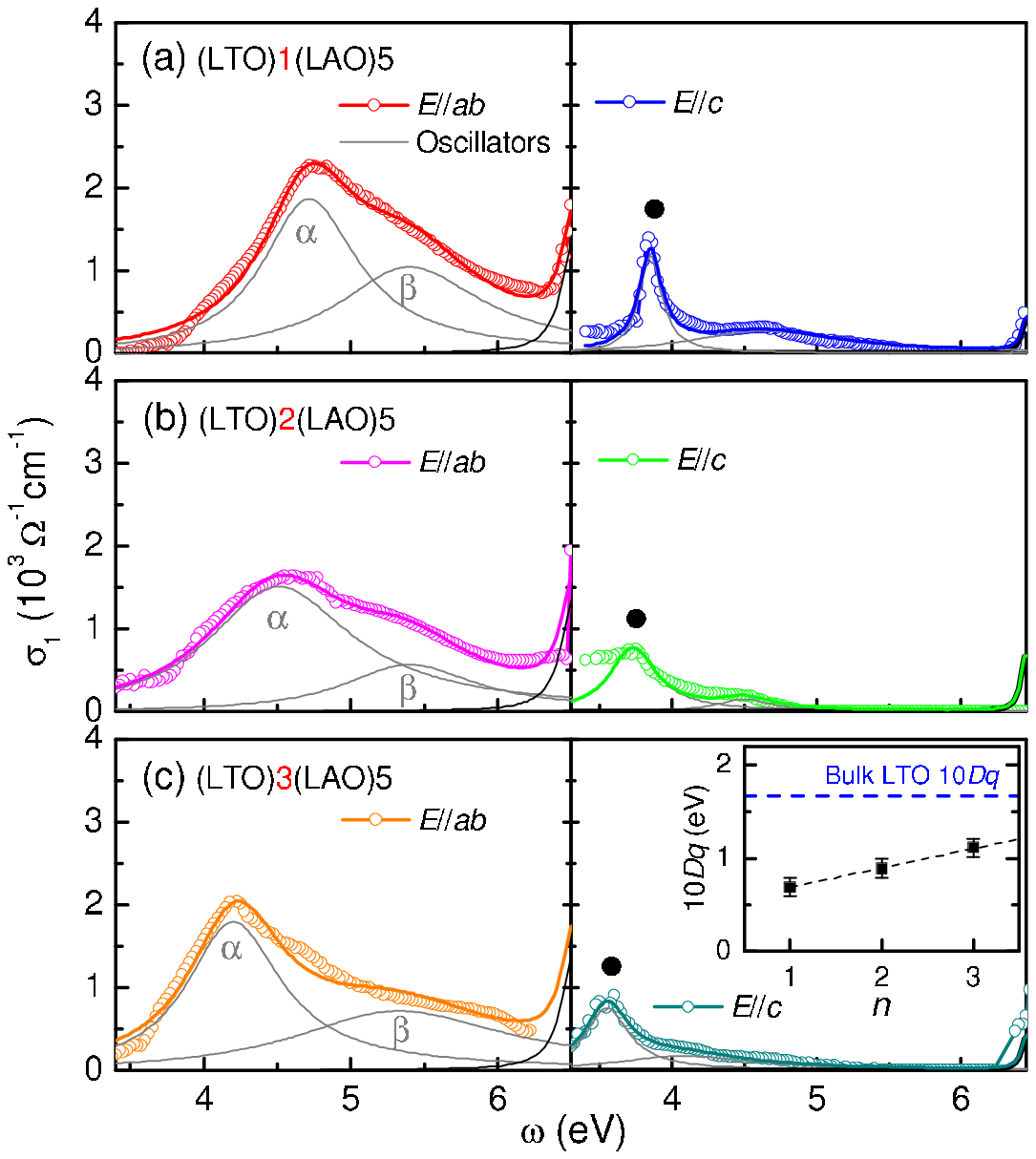}
\caption{\label{fig2} (color online) In-plane (left) and
out-of-plane (right) components of the optical conductivity
spectra of the (LTO)$n$(LAO)5 multilayers with (a) $n$=1, (b)
$n$=2, and (c) $n$=3. (Inset of (c)) 10$Dq$ of Ti$^{3+}$ as a
function of $n$. The dashed (blue) line gives the value in bulk
LTO \cite{Higuchi}.}
\end {figure}

\begin{figure}
\includegraphics[width=3.2 in,bb=130 282 444 616]{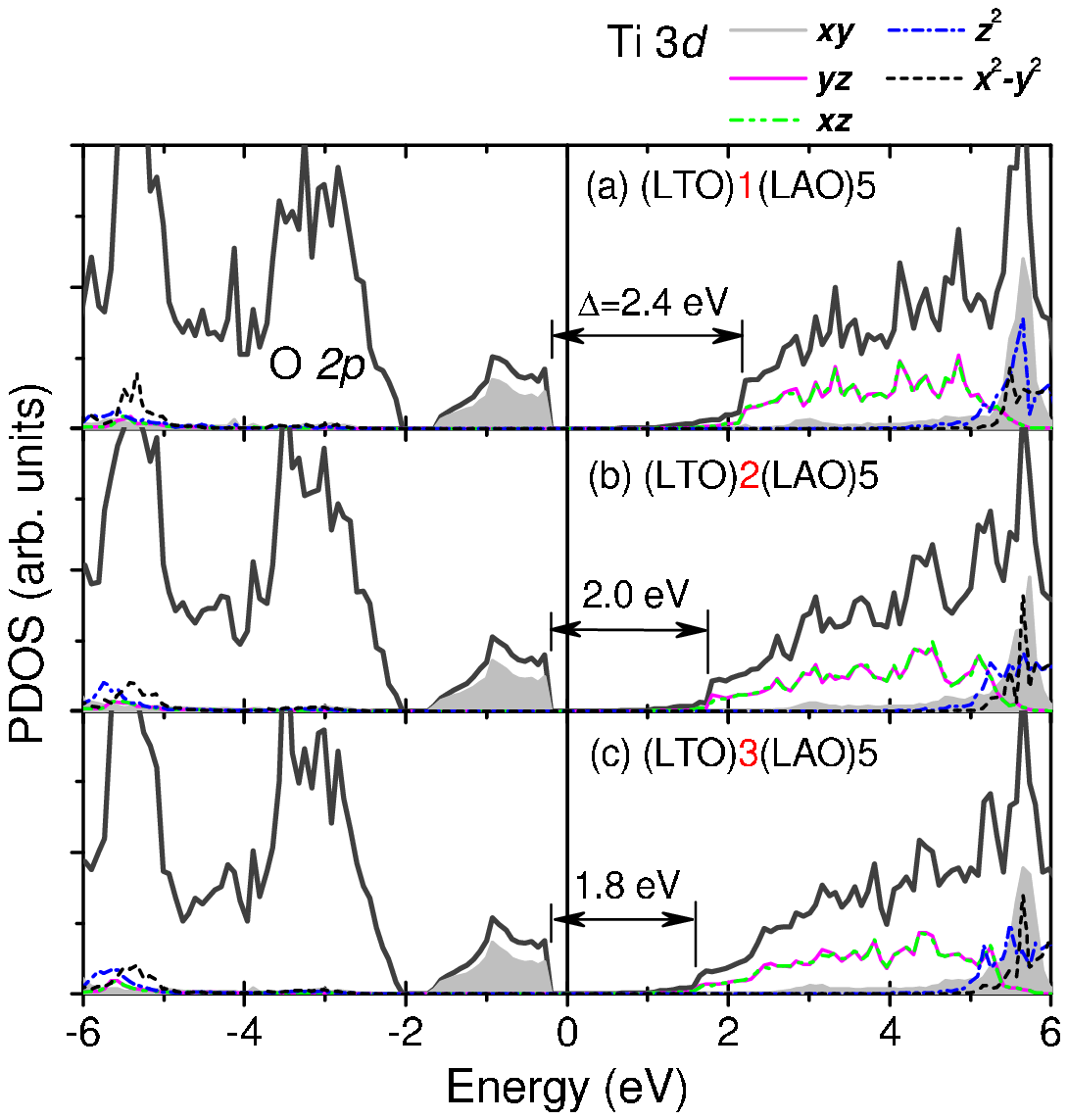}
\caption{\label{fig3} (color online) The partial density of states
of the TiO$_2$ slabs in (a) (LTO)1(LAO)5, (b) (LTO)2(LAO), and (c)
(LTO)3(LAO)5 multilayers calculated by LDA+$U$. $\Delta$ indicates
the energy gap between the occupied $d_{xy}$ orbital and the
unoccupied $d_{yz,zx}$ states.}
\end {figure}

\begin{figure}
\includegraphics[width=3.1 in,bb=145 272 467 586]{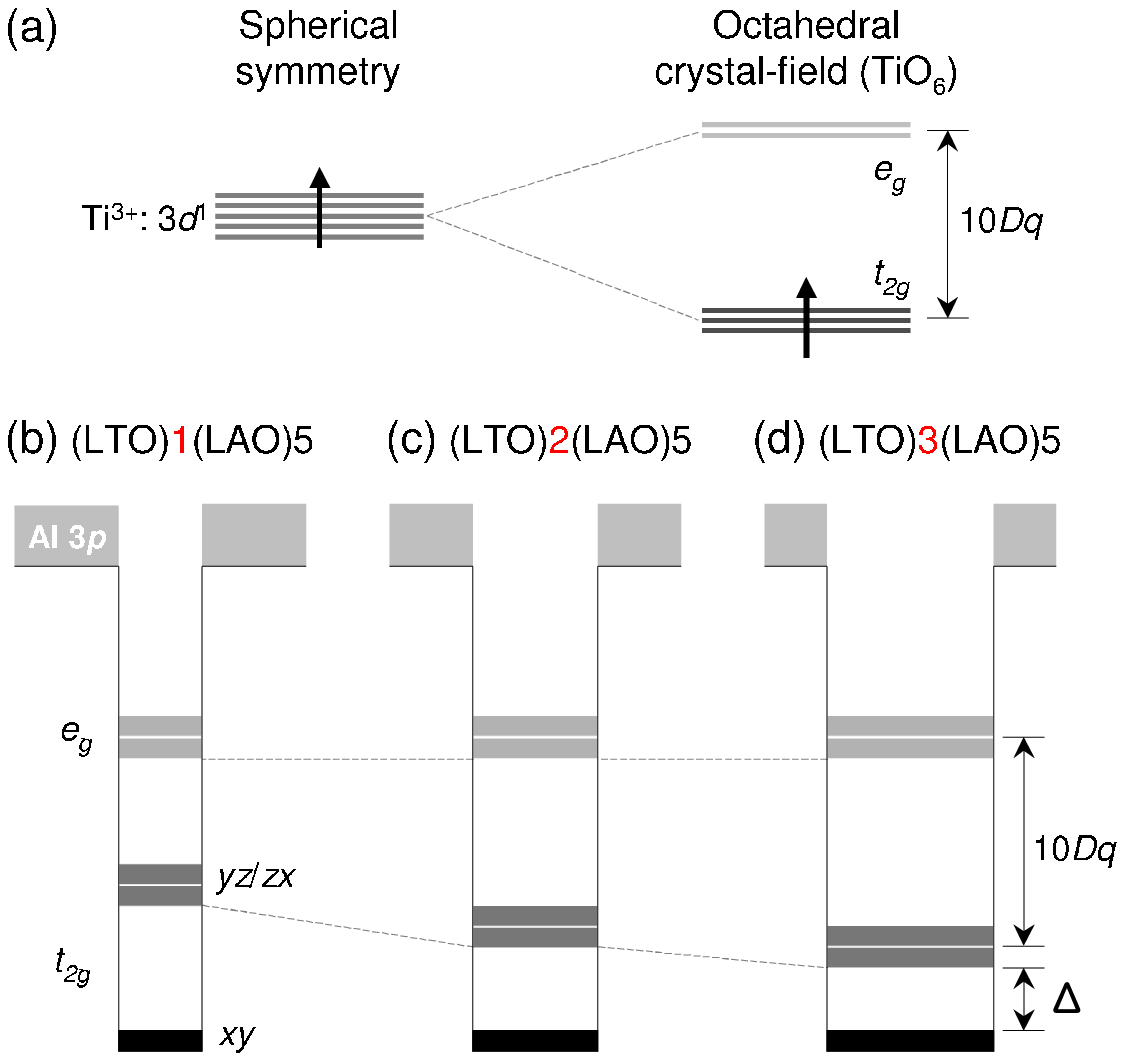}
\caption{\label{fig4} (color online) (a) Crystal field splitting
of a Ti 3$d$ state in octahedral symmetry. Schematic diagram of
additional energy splitting ($\Delta$) of the $t_{2g}$ state in
the multilayers of (b) (LTO)1(LAO)5, (c) (LTO)2(LAO), and (d)
(LTO)3(LAO)5 due to the 2D confinement.}
\end {figure}


\begin{thebibliography}{}
\bibitem{O-La} Another charge transfer transition from O 2$p$ to La 5$d$ would be located above 5.6 eV (Ref. \cite{SGLim}), which is beyond our measurement range.
\end{thebibliography}
\end{document}